\documentclass[twocolumn,showpacs,preprintnumbers,amsmath,amssymb]{revtex4}
\usepackage{graphicx}% Include figure files
\usepackage{dcolumn}% Align table columns on decimal point
\usepackage{bm}% bold math

\begin{document}

\preprint{APS/123-QED}

\title{Nonlinear Lattices Generated from Harmonic Lattices with Geometric Constraints}

\author{S. Takeno$^1$, S. V. Dmitriev$^{2,3}$,  P. G. Kevrekidis$^4$
and A. R. Bishop $^5$}
% \altaffiliation[Also at ]{Physics Department, XYZ University.}
\affiliation{
$^1$ Graduate School, Nagasaki Institute of Applied Science, Nagasaki 851-0193, Japan \\
$^2$ Institute of Industrial Science, the University of Tokyo, 4-6-1 Komaba, Meguro-ku, Tokyo 153-8505, Japan \\
$^3$ National Institute of Materials Science, 1-2-1 Sengen, Tsukuba, Ibaraki 305-0047, Japan \\
$^4$ Department of Mathematics and Statistics, University of Massachusetts \\
Lederle Graduate Research Tower, Amherst, MA 01003-4515, USA \\
$^5$ Center for Nonlinear Studies and Theoretical Division, Los
Alamos National Laboratory, Los Alamos, NM 87545 USA }
\date{\today}

\begin{abstract}
Geometrical constraints imposed  on higher dimensional harmonic
lattices generally lead to nonlinear dynamical lattice models.
Helical lattices obtained by such a procedure are shown to be
described by sine- plus linear-lattice equations. The interplay
between sinusoidal and quadratic potential terms in such models is
shown to  yield localized nonlinear modes identified as intrinsic
resonant modes.
\end{abstract}


\maketitle

\section{ Introduction }

Recent developments in the
physics of intrinsic localized modes (ILM) \cite{1,2}
have shown that the ILMs are not only conceptually important in theoretical
and mathematical physics as ubiquitous fundamental modes in nonlinear,
discrete physical systems \cite{3} but also possess innovative potential
applications in fundamental science and technology in addition to their
experimental observation \cite{4,5,5a}. Areas of such applications cover
coupled Josephson junctions, photonic crystals, optical lattices in
Bose-Einstein condensates, all-optical
logic and switching devices, targeted breaking of chemical bonds,
and so on \cite{6}.

Very recently, the interplay between nonlinear dynamics and geometry
has attracted particular attention in ILM problems \cite{7}. Historically,
such an observation was first made when formulating model Hamiltonians
for the dynamics of bases in DNA  to take care of helical
structure, to the best of the authors' knowledge \cite{8,9}.
%Introduced there
The latter problem gave rise to the intuitive introduction
of the so-called sine-lattice (SL) model \cite{10} in which inter-site
interactions
along a given strand are taken to be sinusoidal
rather than the conventional quadratic ones.
The SL model later turned out to yield a novel type of ILM,
referred to as an intrinsic rotating mode (IRM), also termed a
roto-breather \cite{11}. An example of transition from oscillation to
such rotation
modes in simulations of molecular crystals can be
found in \cite{Wojciechowski}.
Another example of this type is given by curved or bent chains \cite{12}
and long-range interactions on a fixed curved substrate \cite{13}.

Geometric constraints naturally appear in crystalline bodies consisting of
relatively rigid atomic clusters. An important example of this class
of materials are the polymorphs of silica (SiO$_2$),
where the structural units,
i.e., SiO$_4$ tetrahedra are corner-linked by oxygen atoms, and the energy cost
of deformation of the tetrahedra is much greater than the cost of their mutual
rotations. Atoms in the almost rigid clusters move as if they were subjected to
a geometrical constraint. To describe the position of a
finite-size molecular cluster,
one has to introduce not only translational but also rotational degrees of
freedom. It has been demonstrated that the rotational degrees of freedom
can be responsible for the incommensurate
phase in quartz \cite{15,16,17}, negative Poisson's ratio of cristobalite
and quartz \cite{18,19,20,21}, and negative thermal expansion of beta-quartz
\cite{22}. Similar effects can be observed in other materials with microscopic
rotations, such as perovskites (e.g., SrTiO$_3$) containing corner-linked
TiO$_6$ octahedra, the KH$_2$PO$_4$ (KDP) family of crystals with
comparatively rigid PO$_4$ tetrahedra, among others.

Motivated by these investigations, we proposed in a previous paper a
mechanical model in which a set of %otherwise linearly arranged
masses on a linear chain
are re-arranged to slide on fixed rings \cite{14}. Such a model was shown
to contain rich nonlinear dynamics, exhibiting various types of nonlinear
modes ranging from kinks to IRMs. The method employed there
amounts  to applying a specific geometrical constraint to a purely harmonic
lattice, leading eventually to equations having the form of a extended
version of the SL equations.

It is our purpose here to propose a method for studying
the above-mentioned interplay between nonlinearity and geometry
in a more general and transparent way. The basic point of our method
is to take three-dimensional (3D) harmonic lattices as a starting point
on which geometrical constraints are imposed. Since linear systems
constitute a basis for studying physics and mathematics
in general, we expect the present method will give much more insight
into the problem. The work is composed of two parts: the first presents
a general scheme of geometrical constraints, while the second complements
it with a
detailed study of the ILMs in sine- plus linear- lattices resulting from the
helical constraints that we impose on the original 3D harmonic lattice.

This paper is organized as follows: in the next section, we present
a simple 3D harmonic lattice model and consider a general
scheme of the geometrical constraint applied to it.
In Section
III, we introduce the helical constraints as an application of the general
method to arrive at helical lattices described by sine-linear-lattice (SLL)
equations interpolating between SL and linear lattice. Section IV is devoted
to the study of some properties of the SLL equations. Generalization
of the SLL model is made in Section V to include the effect
of an on-site potential, giving rise to the ILMs.
The last section is devoted to concluding remarks.

\section{The linear lattice model and a geometric constraint}

\vspace{0.3cm}
We consider a three-dimensional lattice governed by the Lagrangian $L$
\begin{eqnarray}
L=T-V \equiv \sum_n\frac{m_n}{2}{\left(\dot{x}_n^2+\dot{y}_n^2+\dot{z}_n^2 \right)} \nonumber \\
-\frac{K}{2}\sum_n{\left[(x_{n+1}-x_n)^2+(y_{n+1}-y_n)^2+(z_{n+1}-z_n)^2 \right]},
\label{Lagrangian}
\end{eqnarray}
where $T$ and $U$ are the kinetic energy and the potential energy
of the system, respectively. The quantities ${\displaystyle x_n,y_n,z_n}$
are the dynamical variables associated with the $n$th atom of atomic
mass ${\displaystyle m_n}$. Then, the Euler-Lagrange equations assume
the form
\begin{eqnarray}
m_n\frac{d^2x_n}{dt^2}=K{\left(x_{n+1}+x_{n-1}-2x_n \right)}, \nonumber \\
m_n\frac{d^2y_n}{dt^2}=K{\left(y_{n+1}+y_{n-1}-2y_n \right)}, \nonumber \\
m_n\frac{d^2z_n}{dt^2}=K{\left(z_{n+1}+z_{n-1}-2z_n \right)}.
\label{EqMotion}
\end{eqnarray}
Physically, Eqs. (\ref{EqMotion}) are equations of motion for harmonic lattice
vibrations of a simplified version of a simple cubic lattice,
%with noncentral force
%constants neglected at all,
in which ${\displaystyle x_n,y_n, z_n}$
represent the $x-, y-, z-$ components of the displacement vector
${\displaystyle \vec{r}_n}$ of the $n$th atom from its equilibrium position.

%\section{ Geometrical constraint of the harmonic lattice }

Suppose now that there exists a single variable $s$
%by which the dynamical
%variables $x_n,y_n,z_n$ are given by
such that
\begin{eqnarray}
x_n=f(s_n),\,\,\,\,\,   y_n=g(s_n),\,\,\,\,\,   z_n=h(s_n),
\label{Constraint}
\end{eqnarray}
where $f(s), g(s)$ and $h(s) $ are functions of $s$. The Lagrangian of the system is then written as
\begin{eqnarray}
L=\frac{1}{2}\sum_n m_n{\left[f^{\prime}(s_n)^2+g^{\prime}(s_n)^2+h^{\prime}(s_n)^2\right]}\dot{s}^2  \nonumber \\
-\frac{K}{2}\sum_n{}\big([f(s_{n+1})-f(s_n)]^2  \nonumber \\
+[g(s_{n+1})-g(s_n)]^2+[h(s_{n+1})-h(s_n)]^2\big).
\label{NewLagrangian}
\end{eqnarray}
Then, the Euler-Lagrange equations takes the form
\begin{eqnarray}
m_n{\left[f^{\prime}(s_n)^2+g^{\prime}(s_n)^2+h^{\prime}(s_n)^2\right]}\ddot{s}_n  \nonumber \\
+\frac{m_n}{2}{\left(\frac{d}{ds}[f^{\prime}(s_n)^2+g^{\prime}(s_n)^2+h^{\prime}(s_n)^2]\right)}\dot{s}_n^2 \nonumber \\
=K{\left[f(s_{n+1})+f(s_{n-1})-2f(s_n)\right]}f^{\prime}(s_n)  \nonumber \\
+K{\left[g(s_{n+1})+g(s_{n-1})-2g(s_n)\right]}g^{\prime}(s_n)  \nonumber \\
+K{\left[h(s_{n+1})+h(s_{n-1})-2h(s_n)\right]}h^{\prime}(s_n),
\label{NewEqMotion}
\end{eqnarray}
where ${\displaystyle A^{\prime}(s) \equiv (d/ds)A(s)}$ with $A=f,g,h$.
It can hence be seen from Eqs. (\ref{NewLagrangian}) and (\ref{NewEqMotion})
that the geometric constraint of Eqs. (\ref{Constraint}),
transforms the original linear lattice equations into nonlinear ones.
Thus, depending on the choice of $f,g$ and $h$, we can obtain various
nonlinear equations from the linear lattice model through the one-parameter
geometric constraint.
Equations (\ref{Constraint}) can be generalized to the case in which, e.g.,
the quantities ${\displaystyle x_n, y_n}$ are parametrized by two
variables, ${\displaystyle (s_{1n},s_{2n})}$.
%, or ${\displaystyle (s_{1n},s_{2n},s_{3n})}$.

\section{ Helical Constraint and the sine-linear lattice}

We illustrate the above method by considering situations in which
the constraint functions $f, g, h$ are of the form
\begin{eqnarray}
f(s_n)=a_n\cos(cs_n) \equiv a_n\cos(\theta_n),  \nonumber \\
g(s_n)=a_n\sin(cs_n)\equiv a_n\sin(\theta_n),  \nonumber \\
z_n=v_n s_n \equiv b_n\theta_n,
\label{HelicalConstraint}
\end{eqnarray}
where ${\displaystyle c, a_n, v_n}$ are constants depending on the
site index $n$ and ${\displaystyle b_n=v_n/c}$. Inserting Eqs. (\ref{HelicalConstraint})
into Eq. (\ref{NewLagrangian}), we obtain the Lagrangian of the system in the form
\begin{eqnarray}
L=T-U=\sum_n {\left[\frac{m_n}{2}(a_n^2+b_n^2)\dot{\theta}_n^2\right]}  \nonumber \\
-\sum_n\frac{K}{2}\large[a_{n+1}^2+a_n^2-2a_{n+1}a_n\cos(\theta_{n+1}-\theta_n)  \nonumber \\
+(b_{n+1}\theta_{n+1}-b_n\theta_n)^2\large].
\label{HelicalLagrangian}
\end{eqnarray}
Then, the Euler-Lagrange equations are given by
\begin{eqnarray}
\ddot{\theta}_n=C_n\large[a_{n+1}a_n\sin(\theta_{n+1}-\theta_n) \nonumber \\
-a_na_{n-1}\sin(\theta_n-\theta_{n-1})\large]  \nonumber \\
+C_n{\left[b_{n+1}b_n\theta_{n+1}+b_nb_{n-1}\theta_{n-1}-2b_n^2\theta_n\right]},
\label{HelicalEqMotion}
\end{eqnarray}
where
\begin{eqnarray}
C_n=\frac{K}{m_n(a_n^2+b_n^2)}.
\end{eqnarray}
Equations (\ref{HelicalEqMotion}) interpolate between a generalized version of the
linear lattice
\begin{eqnarray}
\ddot{\theta}_n=C_n\left[b_{n+1}b_n\theta_{n+1}+b_nb_{n-1}\theta_{n-1}-2b_n^2\theta_n\right],
\label{LinearLattice}
\end{eqnarray}
 and that of the sine-lattice equation \cite{9,10}
\begin{eqnarray}
\ddot{\theta}_n=C_n\large[a_{n+1}a_n\sin(\theta_{n+1}-\theta_n)  \nonumber \\
-a_na_{n-1}\sin(\theta_n-\theta_{n-1})\large],
\label{SineLattice}
\end{eqnarray}
where the coefficients of the right-hand side are all site-dependent.
Eq. (\ref{HelicalEqMotion})
are referred to as sine-linear-lattice (SLL) equations.

\section{ A perfect helical lattice}

Here we restrict ourselves to the simplest situations in which
all the coefficients are site-independent. Then, dropping
the subscript $n$ attached to the coefficients in Eq. (\ref{HelicalEqMotion}) leads to
\begin{eqnarray}
\ddot{\theta}_n=J\big[\sin(\theta_{n+1}-\theta_n)-\sin(\theta_n-\theta_{n-1}) \nonumber \\
+\epsilon{\left(\theta_{n+1}+\theta_{n-1}-2\theta_n\right)}\big],
\label{SineLinearLattice}
\end{eqnarray}
where
\begin{eqnarray}
J=Ca^2, \,\,\,\,\,\,   \epsilon=b^2/a^2.
\end{eqnarray}
The non-integrable equation (\ref{SineLinearLattice}) can be considered
as relevant to helical-lattices in which the helicity is determined
by the factor $\epsilon$. Alternatively, Eqs. (\ref{SineLinearLattice}) may be
considered as a modified version of the discrete sine-Gordon equation \cite{23} in which
the conventional on-site term ${\displaystyle \sin(\theta_n)}$ is replaced by
the first term in the right hand side of Eq. (\ref{SineLinearLattice}).

\subsection{Steady state structures}

Equation (\ref{SineLinearLattice}) can be rewritten in the form
\begin{eqnarray}
\ddot{\theta}_n=J(D_{n}-D_{n-1}),\nonumber \\
D_{n}=\sin(\delta_{n})+\epsilon\delta_{n}, \,\,\,\,\,  \delta_{n}=\theta_{n+1}-\theta_n.
\label{SLL}
\end{eqnarray}
Equilibrium solutions to Eq. (\ref{SLL}) can be found from
$D_{n}=D_{n-1}$, which is particularly satisfied when
\begin{eqnarray}
D_{n}=0,
\label{EquilibriumCondition}
\end{eqnarray}
for all $n$. Stable equilibria are subjected to an additional condition,
$D_n^{\prime}(\delta_{n})=\cos(\delta_{n})+\epsilon>0$.

\begin{figure}
\includegraphics{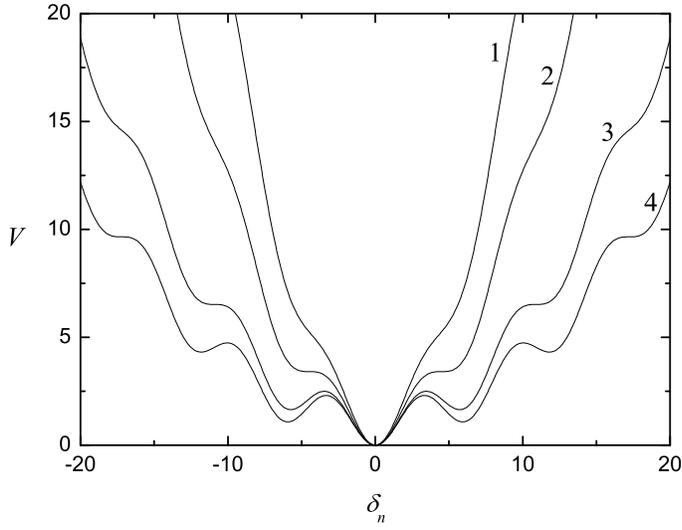}
\caption{The potential $V(\delta_n)$ of Eq.
(\ref{PotentialFunction}) is shown as a function of $\delta_n$ for
different magnitudes of $\epsilon$. Curves 1 to 4 correspond to
$\epsilon=0.4$, $\epsilon_1 \approx 0.2172$, $\epsilon_2 \approx
0.09133$, and $\epsilon_3 \approx 0.05797$. The magnitudes
$\epsilon_l$, at which $l$th minimum of the potential function
appears, are given by Eq. (\ref{NewMinima}). } \label{dfig1}
\end{figure}

To study the physical properties of Eq. (\ref{SineLinearLattice})
which represents
the competition between the linearity and the helicity, let us
consider the potential
function $V(\delta_n)$ associated with $D_n(\delta_{n})$,
\begin{eqnarray}
V(\delta_n)=1-\cos(\delta_n)+\epsilon \frac{\delta_n^2}{2}, \,\,\,\,\,\,   \epsilon >0.
\label{PotentialFunction}
\end{eqnarray}
We pay particular attention to its minimum points. For sufficiently large
$\epsilon$ (weak helicity), the potential $V$
has only one minimum at $\delta_n=0$
and the only stable state of the SLL is the ground state, $\theta_n=const$.
However, with decrease of $\epsilon$, new minima
appear in the potential $V(\delta_n)$, the $l$th minimum appearing at
\begin{eqnarray}
\epsilon_l \approx (4l-1)\frac{\pi}{2}-\sqrt{(4l-1)^2\frac{\pi^2}{4}-2}, \,\,\,\,\, l>0.
\label{NewMinima}
\end{eqnarray}
For example, $\epsilon_1 \approx 0.2172$, $\epsilon_2 \approx 0.09133$,
$\epsilon_3 \approx 0.05797$.
The potential $V(\delta_n)$ is shown in
Fig. \ref{dfig1}  for these values of $\epsilon$ and
also for a larger magnitude, $\epsilon=0.4$, when there is only one minimum.
In the limiting case $\epsilon \rightarrow 0$,
the SLL Eq. (\ref{SineLinearLattice}) degenerates to the SL equation
with energetically indistinguishable minima situated at $\delta_n=2\pi l$.

As the number of minima on the potential $V(\delta_n)$ increases, the variety
of the stable structures of the SLL Eq. (\ref{SineLinearLattice}) also increases.
For $\epsilon_{l+1}<\epsilon<\epsilon_l$, $\delta_{n}$ can obtain $2(l-1)$ values
(positive or negative) in addition to $\delta_{n}=0$,
which is allowed for any $\epsilon$.

\begin{figure}
\includegraphics{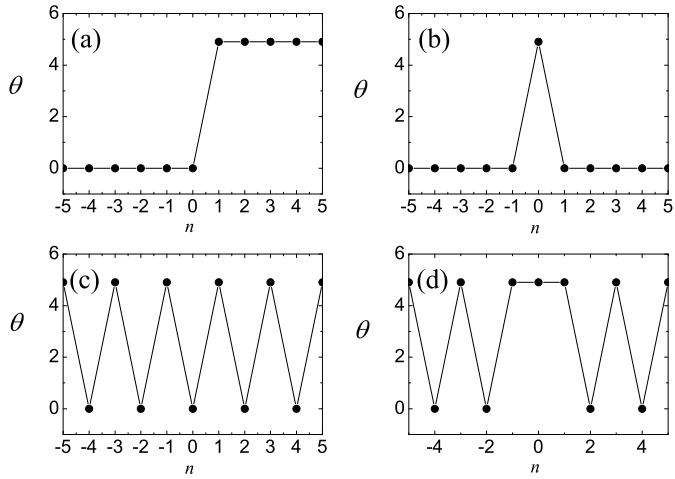}
\caption{Examples of stable equilibrium structures in SLL with
$J=1$, $\epsilon =0.2$. (a) Kink-like structure, (b) point defect
in the ground state structure, (c) zigzag structure, (d) point
defect in the zigzag structure. } \label{dfig2}
\end{figure}

For the particular class of equilibrium solutions expressed by
Eq. (\ref{EquilibriumCondition}), configuration of the $n$th bond,
$\delta_{n}=\theta_{n+1}-\theta_n$, is not affected by the adjacent bonds,
and one can easily create various compacton-like localized defects.
For example, for $\epsilon=0.2$,
$V(\delta_n)$ has minima at $\delta_n=0$ and $\delta_n=\pm 4.9063$.
In Fig. \ref{dfig2} we show several stable configurations of SLL at $\epsilon=0.2$
and $J=1$. Shown are (a) a kink-type structure, (b) a
point defect
in a ground state structure, (c) a zigzag periodic structure, and
(d) a point defect
in the zigzag structure.
The defects in periodic structures presented in
Fig. \ref{dfig2} can be regarded as discrete
compactons \cite{comp} because they are sharply localized and
do not have exponential tails.

\subsection{Perturbation Induced Dynamical Structures}

\begin{figure}
\includegraphics{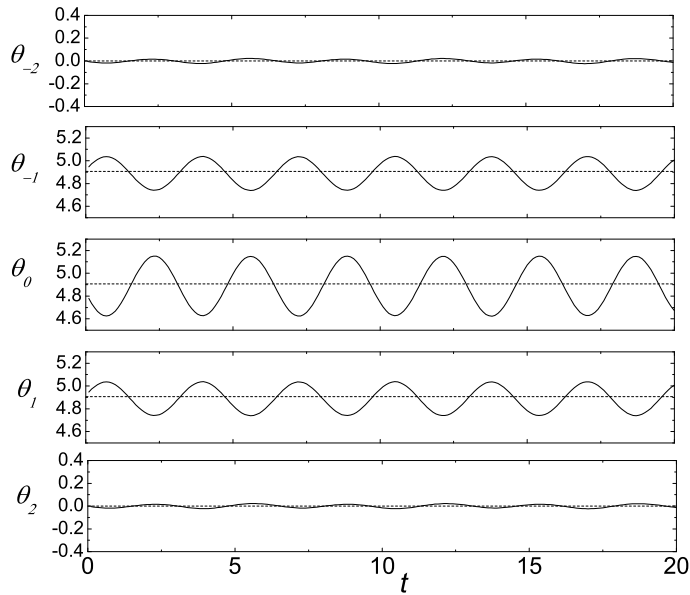}
\caption{Vibration mode excited for zigzag metastable structure
with a defect present in Fig. \ref{dfig2}(d). The mode is
localized on (practically only) three particles. $J=1$, $\epsilon
=0.2$. } \label{dfig3}
\end{figure}

A vibration mode can be excited on the defect in zigzag structure
presented in Fig. \ref{dfig2}(d). The mode is
(practically) localized on three particles
and it is presented in Fig. \ref{dfig3}. The existence of this mode can be
easily
understood.
%The linear vibration spectrum of the zigzag structure has smaller
%maximum frequency than that of the ground state structure.
The 0th particle in
Fig. \ref{dfig2}(d) oscillates with a
frequency close to the upper edge of the
ground state spectrum
which, however, is outside of the frequency band of the zigzag
structure.

%\subsection{ Moving kinks }

\begin{figure}
\includegraphics{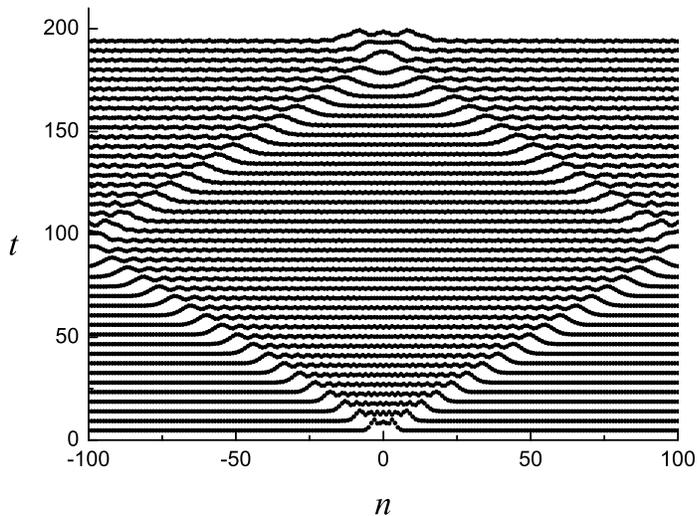}
\caption{Moving kinks initiated by the initial condition
$\theta_{0}(0)=0$, $\dot{\theta}_{0}(0)=3$, and zero initial
conditions for all other particles, presented by the set of plots
$\theta$ vs. $n$ for different times. The height of the kinks is
$|\varphi_1-\varphi_2| \approx 0.9\pi$ and it increases with
increase in initial velocity of $0$th particle. Periodic boundary
conditions are used here to demonstrate the kinks' robustness
against collisions. $J=1$, $\epsilon=0.2$. } \label{dfig4}
\end{figure}

Due to  translational invariance,
the SLL model of Eq. (\ref{SineLinearLattice})
admits solutions with $\theta_n=\varphi$ for arbitrary $\varphi =const$.
Domain walls separating
two domains with $\varphi_1 \neq \varphi_2$ can also be formed.
Here we consider the case when
the difference $|\varphi_1-\varphi_2|$ is not equal to the distance between
the
minima of the potential $V(\delta_n)$ Eq. (\ref{PotentialFunction})
%.
%In this situation, an equilibrium transition between two solutions with
%$\varphi_1 \neq \varphi_2$ is not possible and if it was artificially created,
and which can give rise to moving kinks.
There are many possibilities to initiate such moving kink-like
structures; one is presented in
Fig. \ref{dfig4} for $J=1$, $\epsilon=0.2$. Here the kinks were initiated
by applying initial conditions $\theta_{0}(0)=0$, $\dot{\theta}_{0}(0)=3$,
with zero initial conditions for all other particles.
The height of the kinks is
$|\varphi_1-\varphi_2| \approx 0.9\pi$ and it increases by increasing the
 initial velocity of the $0$th particle. Here we use periodic boundary
conditions to demonstrate that the kinks survive  collisions and hence
are relatively robust objects.

\subsection{ An approximate analytical solution: nonlinear resonant modes (NRM) }

Let us consider situation in which an atom
located at the site $0$ makes a large excursion around one of the
minimum points $\alpha$ of the potential, while the others only perform small
amplitude oscillatory motion. This means that
\begin{eqnarray}
|\theta_0| >> |\theta_n| \,\,\,\,\, {\rm for} \,\,\,\,\, |n| \geq 1.
\label{E16}
\end{eqnarray}
As we expect a highly localized solution, we preserve the full nonlinearity
of the equations for the central site only, while the other equations
are linearized around the equilibrium position. Such a procedure leads to
\begin{eqnarray}
\ddot{\theta}_0=J{\left(-2{\left[\sin(\theta_0)+\epsilon \theta_0 \right]}
+{\left[\cos(\theta_0)+\epsilon \right]}(\theta_1+\theta_{-1})\right)},
\label{E17}
\end{eqnarray}
\begin{eqnarray}
\ddot{\theta}_{\pm 1}=J\big[-{\left[\cos(\theta_0)+1+2\epsilon \right]}\theta_{\pm 1} \nonumber \\
+(1+\epsilon)\theta_{\pm 2}+\sin(\theta_0)+\epsilon \theta_0\big],
\label{E18}
\end{eqnarray}
\begin{eqnarray}
\ddot{\theta}_n=J(1+\epsilon)(\theta_{n+1}+\theta_{n-1}-2\theta_n), \,\,\,\, {\rm for} \,\,|n| \geq 2.
\label{E19}
\end{eqnarray}
We now seek solutions to this set of equations in the form
\begin{eqnarray}
\theta_0=\alpha+u_0,\,\,\,\,
{\rm with} \,\,\,\,\sin(\alpha)+\epsilon \alpha=0,  \nonumber \\
\theta_n \equiv u_n, \,\,\, {\rm for} \,\,\, |n| \geq 1.
\label{E20}
\end{eqnarray}
When, in Eq. (\ref{E20}), $\alpha =0$, all particles oscillate near one
well of the potential $V(\delta_n)$, while for $\alpha \neq 0$ the $0$th particle
oscillates in a different well.

Inserting Eq. (\ref{E20}) into Eqs. (\ref{E17}), (\ref{E18}) and (\ref{E19})
and retaining solely terms linear with respect to $u_0$ and $u_n (n \neq 0)$,
we arrive at a set of equations which are of the same form as those appearing
in one-impurity problems in harmonic lattice vibrations. Thus, setting
\begin{eqnarray}
u_n=v_n\exp(-i\omega t),
\label{E21}
\end{eqnarray}
where ${\displaystyle v_n}$ is time-independent and $\omega$ is a constant,
we reduce the above set of equations to the form
\begin{eqnarray}
(L^{(0)}-\omega^2)v_n=L^{\prime}v_n, \nonumber \\
{\rm with} \,\,\,\,  L^{(0)}v_n=J(1+\epsilon)(2v_n-v_{n+1}-v_{n-1}),
\label{E22}
\end{eqnarray}
with
\begin{eqnarray}
L^{\prime}v_0=J{\left[1-\cos(\alpha)\right]}(2v_0-v_1-v_{-1}), \nonumber \\
L^{\prime}v_{\pm 1}=J{\left[1-\cos(\alpha)\right]}(v_{\pm 1}-v_0), \nonumber \\
L^{\prime}v_n=0 \,\,\,\, {\rm for} \,\,\,\, n \neq 0, \pm 1.
\label{E23}
\end{eqnarray}
From the above equations, the ILMs we are seeking are presumed to exist
within the frequency band ${\displaystyle \omega(k)^2=2J(1+\epsilon)[1-\cos(k)]}$
of the perfect harmonic lattice characterized by wavenumber $k$.
Such a nonlinear localized in-band modes are referred to as (intrinsic) nonlinear
resonant modes (NRM). Then, Eqs. (\ref{E22}) and (\ref{E23}) can be handled
by introducing lattice Green's functions
\begin{eqnarray}
g(n) \equiv g(n,\omega^2) \nonumber \\
=\frac{1}{N}\sum_k \frac{\exp(ikn)}{2J(1+\epsilon){\left[1-\cos(k)\right]}-\omega^2-i\gamma}, \nonumber \\
\gamma \rightarrow 0_+.
\label{E24}
\end{eqnarray}
Analytical expressions for the lattice Green's functions $g(n)$
can be obtained by replacing the sum with respect to $k$ by an integral
over the first Brillouin zone as follows
\begin{eqnarray}
g(n)=\frac{1}{2J(1+\epsilon)}\frac{1}{\pi}\int_0^{\pi}\frac{\cos(nx)dx}{y-\cos(x)-i\gamma} \nonumber \\
=\frac{i^{n+1}}{2J(1+\epsilon)}\left[C_n(y)-iS_n(y)\right],
\label{E25}
\end{eqnarray}
where
\begin{eqnarray}
C_n(y)=\int_0^{\infty}\cos(yt)J_n(t)dt, \nonumber \\
S_n(y)=\int_0^{\infty}\sin(yt)J_n(t)dt, \nonumber \\
{\rm with} \,\,\,\, y=1-\frac{\omega^2}{2J(1+\epsilon)},
\label{E26}
\end{eqnarray}
in which ${\displaystyle J_n(t)}$ is the Bessel function of the $n$th order.
We note that the quantity
\begin{eqnarray}
C_0(y)=\frac{1}{\sqrt{1-y^2}}, \nonumber \\
{\rm with} \ \ \ \ S_0(y)=0 \ \  {\rm for} \ \ \ 0<y<1,
\label{E27}
\end{eqnarray}
represents the density of states of the band. % ${\displaystyle \omega(k)^2}$.
In terms of the ${\displaystyle g_n}$s so obtained, Eqs. (\ref{E22}) and (\ref{E23}) are rewritten as
\begin{eqnarray}
u_n=J\left[1-\cos(k)\right]\big[g(n)(2v_0-v_1-v_{-1}) \nonumber \\
+g(n-1)(v_1-v_0)+g(n+1)(v_{-1}-v_0)\big].
\label{E28}
\end{eqnarray}
We pay particular attention to an s-like mode having the symmetry property
\begin{eqnarray}
v_1=v_{-1},
\label{E29}
\end{eqnarray}
and use the identity relation,
\begin{eqnarray}
J(1+\epsilon){\left[2g(n)-g(n+1)-g(n-1)\right]} \nonumber \\
-\omega^2g(n)=\Delta_n,
\label{E30}
\end{eqnarray}
to obtain
\begin{eqnarray}
v_n=\lambda{\left[\omega^2g(n)+\Delta_n\right]}(v_0-v_1),
\label{E31}
\end{eqnarray}
where
\begin{eqnarray}
\lambda=\frac{1-\cos(\alpha)}{1+\epsilon},
\label{E32}
\end{eqnarray}
and ${\displaystyle \Delta_n}$ are Kronecker's delta.

In terms of the dimensionless frequency ${\displaystyle \xi=\omega^2/2J(1+\epsilon)}$,
an equation giving the eigenfrequency of the NRM is obtained
from Eqs. (\ref{E25}), (\ref{E27}), and (\ref{E31}) as follows:
\begin{eqnarray}
1-\lambda-\lambda \xi -i\lambda \xi^2 C_0(y)=0.
\label{E33}
\end{eqnarray}
The dimensionless squared eigenfrequency $\xi$ is therefore obtained as
\begin{eqnarray}
\xi=\frac{1-\lambda}{\lambda}+i\xi_0\sqrt{\frac{\xi_0}{2-\xi_0}}.
\label{E34}
\end{eqnarray}
The localization of the NRM is given by the equation
\begin{eqnarray}
v_n=\lambda \omega^2g(n)(v_0-v_1), \,\,\,\, n \neq 0.
\label{E35}
\end{eqnarray}
As mentioned before, here the localized mode around the local minimum
point of the potential function is of resonant type, the eigenfrequency
which
appears close to bottom of the frequency band
${\displaystyle \omega(k)^2=2J(1+\epsilon){\left[1-\cos(k)\right]}}$.
Apart from the factor $\alpha$, which is the position of the local minimum,
the localization properties of the NRM are essentially different from those of
the ILM in that it exhibits oscillatory slow decay, contrary to
what is the case for the ILM. This point will be examined in detail below.

\begin{figure}
\includegraphics{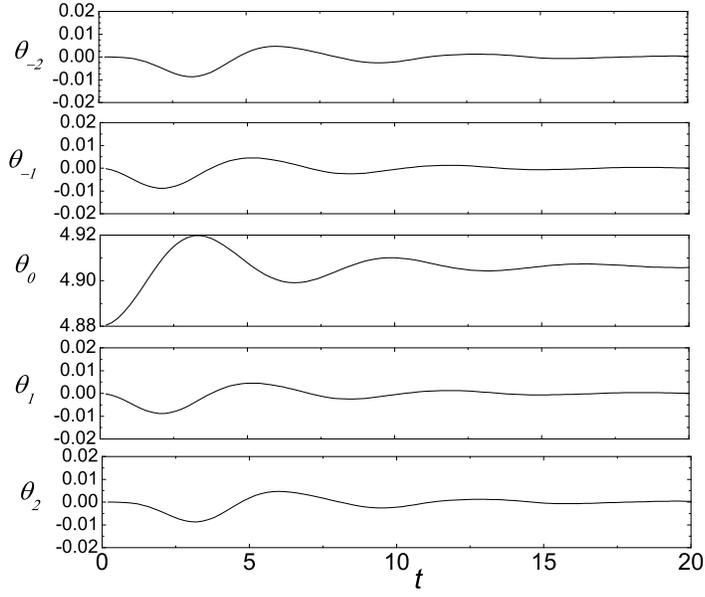}
\caption{An attempt to excite a NRM in a SLL lattice Eqs.
(\ref{SineLinearLattice}) by means of initial conditions
$\theta_{0}(0)=4.88$, $\dot{\theta}_{0}(0)=0$, with zero initial
conditions for all other particles. $J=1$, $\epsilon=0.2$. The
$0$th particle oscillates near the minimum of $V(\delta_n)$
situated at $\delta_n \approx 4.9063$, while other particles
oscillate near the minimum at $\delta_n=0$. Due to the rather
strong interaction with the lattice, the energy of excitation is
``dissipated'' to the lattice after only a few oscillations. }
\label{dfig5}
\end{figure}

We have attempted to excite NRMs in SLL
of Eqs. (\ref{SineLinearLattice}). Such a numerical calculation is presented
in Fig. \ref{dfig5}. Here we set $J=1$, $\epsilon=0.2$ and for the initial
conditions, $\theta_{0}(0)=4.88$, $\dot{\theta}_{0}(0)=0$, with zero initial
conditions for all other particles. For this choice
of $\epsilon$, the potential $V(\delta_n)$ has minima at $\delta_n =0$ and
$\delta_n \approx \pm 4.9063$. One can see that the $0$th particle oscillates
near the minimum of $V(\delta_n)$ situated at $\delta_n \approx 4.9063$,
while the remaining particles oscillate near the minimum at $\delta_n=0$.
Due to the rather strong interaction with the linear excitations of
the lattice, the energy of the NRM is rapidly imparted to the lattice
(after a few oscillations).
The lifetime of the excitation does not increase much, either by changing
the initial deviation of the $0$th particle from the potential minimum or
by changing $\varepsilon$.

\section{ Inclusion of the on-site potential }

\subsection{ Derivation of the model }

It is of interest to see what happens if we include on-site potentials in our
original lattice.
This is done by generalizing Eq. (\ref{Lagrangian}) in the form
\begin{eqnarray}
L=\sum_n\frac{m_n}{2}{\left(\dot{x}_n^2+\dot{y}_n^2+\dot{z}_n^2 \right)} \nonumber \\
-\frac{K}{2}\sum_n\big[(x_{n+1}-x_n)^2+(y_{n+1}-y_n)^2        \nonumber \\
+(z_{n+1}-z_n)^2+\frac{\kappa_1}{2}x_n^2+\frac{\kappa_2}{2}y_n^2\big],
\label{LwithOnsite}
\end{eqnarray}
where ${\displaystyle \kappa_1}$ and ${\displaystyle \kappa_2}$ are constants.
Applying the helical constraint, Eq. (\ref{HelicalConstraint}) leads to
\begin{eqnarray}
L=K-U=\sum_n {\left[\frac{m_n}{2}(a_n^2+b_n^2)\dot{\theta}_n^2\right]}  \nonumber \\
-\sum_n\frac{K}{2}\big[a_{n+1}^2+a_n^2-2a_{n+1}a_n\cos(\theta_{n+1}-\theta_n)  \nonumber \\
+(b_{n+1}\theta_{n+1}-b_n\theta_n)^2  \nonumber \\
+\frac{a_n^2(\kappa_2-\kappa_1)}{4}\cos{\left[1-\cos(2\theta_n)\right]}\big].
\label{LwithOnsitewithConstraint}
\end{eqnarray}
It is seen that such a procedure adds a sine-Gordon type on-site potential
${\displaystyle 1-\cos(2\theta_n)}$ to the Lagrangian. The equations of motion
for ${\displaystyle \theta_n}$ then assume the form
\begin{eqnarray}
\ddot{\theta}_n=L_n\big[a_{n+1}a_n\sin(\theta_{n+1}-\theta_n) \nonumber \\
-a_na_{n-1}\sin(\theta_n-\theta_{n-1}) \nonumber \\
-\lambda a_n^2\sin(2\theta_n)\big]  \nonumber \\
+L_n\big[b_{n+1}b_n\theta_{n+1}+b_nb_{n-1}\theta_{n-1} \nonumber \\
-(b_{n+1}b_n+b_nb_{n-1})\theta_n\big],
\label{EMwithOnsite}
\end{eqnarray}
where ${\displaystyle \lambda=(\kappa_2-\kappa_1)/2}$. When all the
coefficients are site-independent, the above equations reduce to
\begin{eqnarray}
\ddot{\theta}_n=J\big[\sin(\theta_{n+1}-\theta_n)-\sin(\theta_n-\theta_{n-1}) \nonumber \\
-\lambda \sin(2\theta_n)+\epsilon{\left(\theta_{n+1}+\theta_{n-1}-2\theta_n\right)}\big].
\label{SLLwithOnsite}
\end{eqnarray}
The presence of the on-site potential here can be speculated
to induce the existence
of localized modes with longer lifetime than that of the case of
Eq. (\ref{SineLinearLattice}).
In the following, we demonstrate the existence of ILMs in the model of
Eq. (\ref{SLLwithOnsite}).

\subsection{ Intrinsic localized modes }

The linear spectrum of the SLL with the
on-site potential, Eq. (\ref{SLLwithOnsite}), is given by
\begin{eqnarray}
\omega^2=J[4(1+\epsilon)\sin^2(k/2)+2\lambda],
\label{SLLwithOnsiteSpectrum}
\end{eqnarray}
with
\begin{eqnarray}
\omega_{\min}=\sqrt{2J\lambda}, \,\,\,\,\,\,
\omega_{\max}=\sqrt{J[4(1+\epsilon)+2\lambda]}.
\label{MinMax}
\end{eqnarray}

When $\omega_{\max}/2 < \omega_{\min}$, i.e.,
$\lambda >(2/3)(1+\epsilon)$, one can look for an ILM
oscillating with frequency $\omega_{\max}/2 <\omega < \omega_{\min}$
so that all higher harmonics $l\omega$, with an integer
$l>1$, lie above $\omega_{\max}$. In this situation the ILM
would not interact with the lattice preserving its identity.

\subsubsection{ ILM with all particles in one potential well }

\begin{figure}
\includegraphics{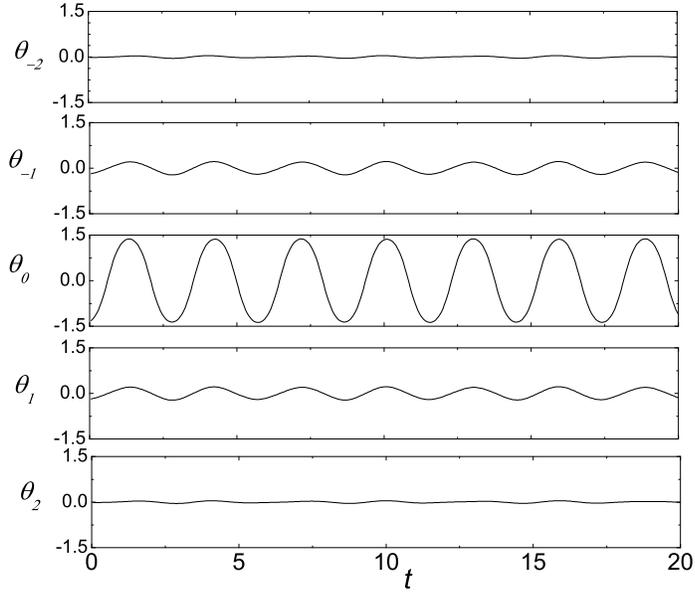}
\caption{ILM with all particles oscillating in one well of the
on-site potential. The mode has amplitude $A_1 \approx 1.37$ and
frequency $\omega \approx 2.15$, which is considerably lower than
the bottom edge of the linear spectrum, $\omega_{\min} \approx
3.162$ with all higher harmonics lying above the upper edge,
$\omega_{\max} \approx 3.847$. $J=1$, $\epsilon=0.2$, and
$\lambda=5$. } \label{dfig6}
\end{figure}

As an example, we take $J=1$, $\epsilon=0.2$, and $\lambda=5$,
with $\omega_{\min}=\sqrt{10} \approx 3.162$, and
$\omega_{\max}=\sqrt{14.8} \approx 3.847$ so that the necessary
condition of the existence of ILM, $\omega_{\max}<2\omega_{\min}$,
is fulfilled. A large-amplitude localized mode can be excited by
choosing the initial
deviation and/or initial velocity of a particle. As an
example, we initialize the $0$th particle with
$\theta_0(0)=1.5$, $\dot{\theta}_0(0)=0$ with zero initial
conditions for other particles and, after some stabilization period,
a steady oscillatory motion is observed (see Fig. \ref{dfig6}).
One can see that the ILM is highly localized. The mode has a frequency
$\omega \approx 2.15$, which is considerably lower than
the bottom edge of the linear spectrum, $\omega_{\min}$,
with all higher harmonics lying above the upper edge,
$\omega_{\max}$.

The ILM can be expressed analytically assuming that
\begin{eqnarray}
\theta_0(t)=A_1\sin(\omega t)+A_3\sin(3\omega t), \nonumber \\
\theta_1(t)=\theta_{-1}(t)=B_1\sin(\omega t),  \nonumber \\
\theta_n(t) \equiv 0  \,\,\,\, {\rm for} \,\,\,\, |n|>1,
\label{ILM1}
\end{eqnarray}
with $A_3 \ll A_1$ and $B_1 \ll A_1$. Parameters of the approximate solution
Eq. (\ref{ILM1}) can be expressed in terms of the ILM amplitude,
$A_1$:
\begin{eqnarray}
\omega^2=J\left[ 2(1+\epsilon+\lambda)-\frac {1}{4}(1+4\lambda)A_1^2  \right],  \nonumber \\
A_3=\frac{(1+4\lambda)A_1^3}{12\left[9\frac{\omega^2}{J}-2(1+\epsilon+\lambda)\right]},  \nonumber \\
B_1=\frac{(1+\epsilon)A_1-\frac{1}{8}A_1^3}{2(1+\epsilon+\lambda)-\frac{\omega^2}{J}}.
\label{ILM1parameters}
\end{eqnarray}

Let us analyze the obtained solution. First of all we note
that the ILM frequency, $\omega$, cannot be greater than
$\omega_{\min}$, i.e., we can only look for a mode with frequency below
$\omega_{\min}$ [see Eq. (\ref{MinMax})], which is possible only for ILMs
with sufficiently large amplitude,
\begin{eqnarray}
A_1>\sqrt{\frac{8(1+\epsilon)}{1+4\lambda}}.
\label{AmplitudeCondition}
\end{eqnarray}
For example, for $\epsilon=0.2$, and $\lambda=5$, one has $A_1>0.68$.
We have confirmed numerically that, for this choice of parameters,
our solution can be used for $A_1 \sim 1$.

\begin{figure}
\includegraphics{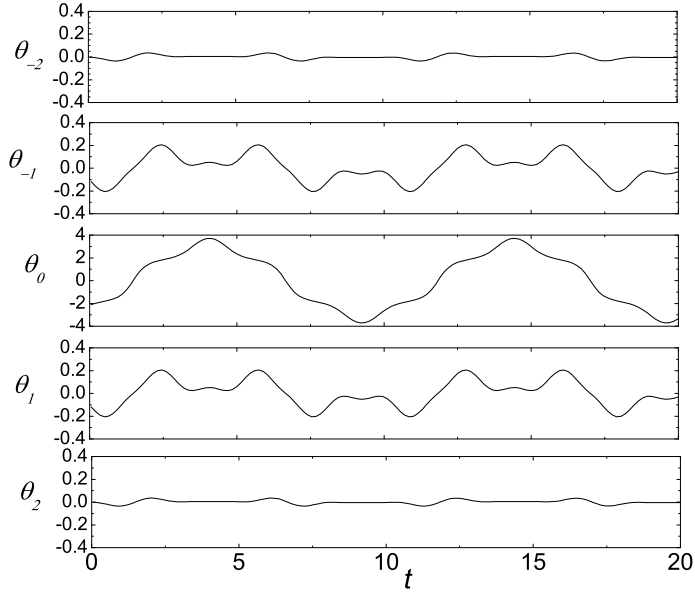}
\caption{Same as in Fig. \ref{dfig6} but for larger amplitude,
$A_1 \approx 3.71$. The ILM has frequency $\omega \approx 0.607$
which is significantly lower than the bottom edge of the linear
spectrum, $\omega_{\min} \approx 3.162$. The mode radiates energy
extremely slowly due to the interaction of higher harmonics with
the linear spectrum. Note the difference in the ordinate scale for
the middle panel. $J=1$, $\epsilon=0.2$, and $\lambda=5$. }
\label{dfig7}
\end{figure}

For larger amplitudes the ILM is highly localized (see Fig. \ref{dfig7})
but the solution Eq. (\ref{ILM1}), (\ref{ILM1parameters}) becomes
inaccurate because it takes into account only cubic anharmonicity.
The ILM in Fig. \ref{dfig7} has amplitude $A_1 \approx 3.71$ and
frequency $\omega \approx 0.607$, which is significantly lower the
bottom edge of the linear spectrum, $\omega_{\min} \approx 3.162$.
The mode radiates energy extremely slowly due to the interaction
of higher harmonics with the linear spectrum.
Note the difference in the ordinate scale for the middle panel.

For small ILM amplitudes, approaching
the allowed limit, Eq. (\ref{AmplitudeCondition}), the solution
profile becomes wider
and our assumption that it is localized on three particles becomes invalid.
Estimation for the limiting amplitude, Eq. (\ref{AmplitudeCondition}), is
also valid only for a highly localized ILM. In fact, it is possible to excite
an ILM with a very small amplitude but the width of such ILM increases
considerably and its frequency approaches $\omega_{\min}$ from below.
The smooth transformation of the ILM from a highly localized (as the amplitude
decreases) into a very broad one
is not surprising because
for smooth solutions, $(\theta_{n+1}-\theta_n) \ll 1$, one has
$\sin(\theta_{n+1}-\theta_n) \approx (\theta_{n+1}-\theta_n)$
and the SLL with the on-site potential, Eq. (\ref{SLLwithOnsite}),
can be approximated by the well-known continuum sine-Gordon equation.
The ILM is, then, nothing but the analog of breather
solution to the sine-Gordon equation. Using the Lorentz invariance of the
sine-Gordon equation one can create a moving small-amplitude
breather.

We also note that, when $\epsilon \gg 1$ and $\lambda \gg 1$, the first
two terms in the right hand side of Eq. (\ref{SLLwithOnsite}) can be
neglected. In this situation, when $\epsilon$ and $\lambda$ are
of the same order of magnitude, our solution, Eqs. (\ref{ILM1}) and
(\ref{ILM1parameters}), describes a highly localized ILM in
the Frenkel-Kontorova model \cite{23}.

\subsubsection{ ILM with one particle in a different potential well }

\begin{figure}
\includegraphics{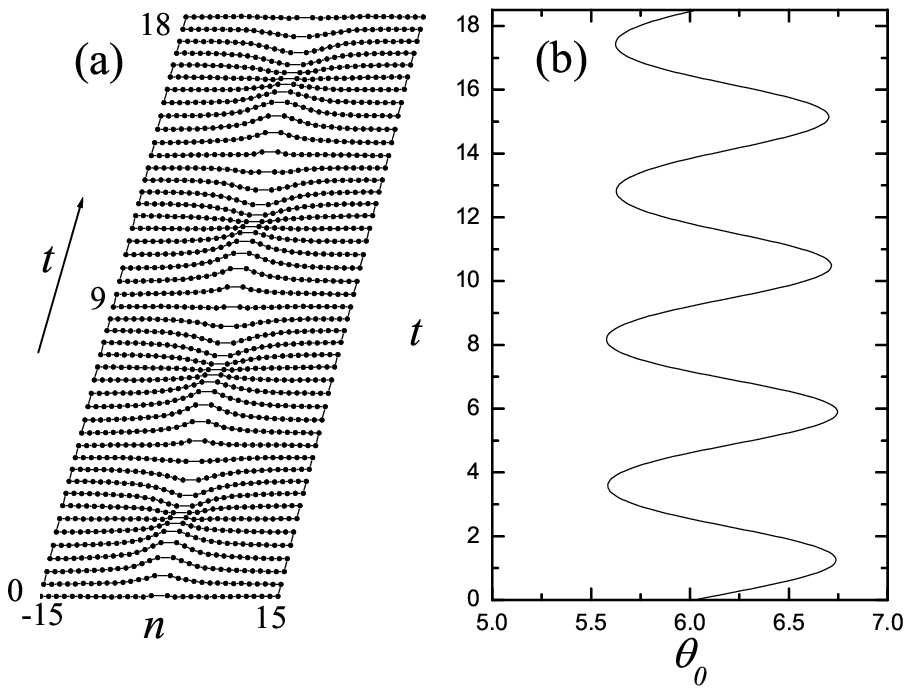}
\caption{ILM with one particle oscillating in a potential well
different from that where other particles are located. (a) Plots
$\theta$ vs. $n$ for different times showing the dynamics of
particles (the $0$th particle is not shown here because its
coordinate differs from coordinates of other particles
significantly, by roughly $2\pi$). (b) Dynamics of $0$th particle.
The ILM is rather broad. The $0$th particle oscillates with the
amplitude $A(\theta_0)\approx 0.6$ while its nearest neighbors
have nearly the same  amplitude, $A(\theta_{\pm 1})\approx 0.52$,
and the amplitudes decrease rather slowly with deviation from
$0$th particle. $J=1$, $\epsilon=0.05$, and $\lambda=1$. }
\label{dfig8}
\end{figure}

The SLL lattice with on-site potential supports localized modes
of another type, when one particle is trapped in a potential well
different from the well occupied by the other particles. To demonstrate
this mode, we take $J=1$, $\epsilon=0.05$, and $\lambda=1$,
with $\omega_{\min}=\sqrt 2 \approx 1.414$, and
$\omega_{\max}=\sqrt{6.2} \approx 2.490$ so that the necessary
condition, $\omega_{\max}<2\omega_{\min}$, is fulfilled.
Setting $\theta_0(0)=2.9$, $\dot{\theta}_0(0)=0$ with zero
initial conditions for other particles, after some time a steady oscillatory
motion presented in  Fig. \ref{dfig8} was observed.
The $0$th particle was found to
oscillate with $\omega \approx 1.35$ which is below
the linear vibration band but all higher harmonics are above the band.
One can see that the ILM is not highly localized. The $0$th particle
oscillates with the amplitude $A(\theta_0)\approx 0.6$ near the
coordinate shifted by roughly $2\pi$ with respect to the neutral
position of all other particles [shown in Fig. \ref{dfig8}(b)]. Its nearest
neighbors oscillate with nearly the
same  amplitude, $A(\theta_{\pm 1})\approx 0.52$,
and the amplitudes decrease rather slowly with deviation from the
$0$th particle [see Fig. \ref{dfig8}(a)].

\section{ Concluding remarks }

In this paper, we have outlined a general scheme to apply geometrical
constraints to a simplified version of the 3D harmonic lattice with or
without an on-site potential to obtain nonlinear dynamical lattice equations.
Depending on the type of the constraints, we arrive at various nonlinear
equations, being interested in the coherent structures that arise in them.

We then studied, more specifically,
a helical constraint by which the original 3D
harmonic lattice was shown to reduce to 1D sine-plus-linear lattice equations
or helical lattice equations. This
formulation revealed several
interesting features:
\begin{enumerate}
\item It provides us with a systematic derivation of
the sine-lattice equations which were derived heuristically
in earlier works.
\item Physically, applying the helical constraint
amounts to transforming the original harmonic lattices (possessing the
hard potential) to a nonlinear dynamical lattice that supports
resonant and multi-kink modes. Seeking other
kinds of nonlinear modes will be an interesting topic for future study.
\item The method developed
here may be relevant to the study of nonlinear excitations in biomolecules
with helical structure such as DNA and proteins.
There are several areas of condensed matter physics where the
existence of a complex energy landscape with
a number of local minima in the potential function is presumed
to play a crucial role in determining their properties. Such examples
can, for instance, be found in the study of
glasses and proteins. Application of the techniques employed
here to these problems would be another topic warranting further
investigation.
\item Finally, another relevant generalization would be to consider
geometric constraints different than the helical one, and derive and
study the
ensuing (reduced) dynamical models.
\end{enumerate}

These topics are currently under
study and will be reported elsewhere.

This work was partially supported by NSF-DMS-0204585, NSF-CAREER, and the
Eppley Foundation for Research (PGK).
Work at Los Alamos is supported by the US DoE.

%*********************
%*** Fig. 1 (Potential)
%*********************

%*********************
%*** Fig. 2 (Stable structures)
%*********************

%*********************
%*** Fig. 3 (Vibration mode on a defect)
%*********************

%*********************
%*** Fig. 4 ( Kinks )
%*********************

%*********************
%*** Fig. 5 ( NRM )
%*********************

%*********************
%*** Fig. 6 (ILM 1)
%*********************

%*********************
%*** Fig. 7 (ILM 1 wide)
%*********************

%*********************
%*** Fig. 8 (ILM 2)
%*********************


\begin{references}

\bibitem{1} S. Takeno, Prog. Theor. Phys. {\bf 75}, 1 (1986);
S. Takeno, K. Kisoda, and A. Sievers, Prog. Theor. Phys. Suppl. {\bf 94}, 242 (1988).
\bibitem{2} A. J. Sievers and S. Takeno, Phys. Rev. Lett. {\bf 61}, 970 (1988).
\bibitem{3} For recent reviews, see for example, S. Flach and C. R. Willis,
Phys. Rep., {\bf 295}, 181 (1998); D. Hennig and G. Tsironis, Physics Reports {\bf 307}, 333 (1999);
P.G. Kevrekidis, K.O. Rasmussen, and A.R. Bishop, Int. J. Mod. Phys. B {\bf 15} (2001) 2833-2900;
J.Ch. Eilbeck and M. Johansson, in {\em Localization and Energy Transfer in Nonlinear Systems},
L. Vazquez, R.S. MacKay, and M.P. Zorzano (eds.), (World Scientific, Singapore, 2003), p.44.
\bibitem{4} U. T. Schwarz, L.Q. English and A.J. Sievers,
Phys. Rev. Lett. {\bf 83}, 223 (1999).
\bibitem{5} N. K. Voulgarakis, G. Kalosakas, A.R. Bishop and
G.P. Tsironis,
Phys. Rev. B {\bf 64}, 020301(R) (2001) and references cited therein.
\bibitem{5a} E. Tr{\'{\i}}as, J.J. Mazo, and T.P. Orlando,
Phys. Rev. Lett. {\bf 84}, 741 (2000); P. Binder, D. Abraimov, A.V. Ustinov, S. Flach, and Y. Zolotaryuk,
Phys. Rev. Lett. {\bf 84}, 745 (2000).
\bibitem{6} For a very recent review on the latest developments, see, e.g.,
D. K. Campbell, S. Flach and Y. Kivshar, Phys. Today {\bf 57}, 43 (2004).
\bibitem{7} See, e.g., P.G. Kevrekidis,
B.A. Malomed and A.R. Bishop,
Phys. Rev. E {\bf 66}, 046621 (2002) and references cited therein.
\bibitem{8} S. Yomosa, Phys. Rev. A {\bf 27}, 2120 (1983); A {\bf 30}, 474 (1984).
\bibitem{9} S. Homma and S. Takeno, Prog. Theor. Phys. {\bf 72}, 679 (1984).
\bibitem{10} S. Takeno and S. Homma J. Phjys. Soc. Jpn. {\bf 65}, 2547 (1986).
\bibitem{11} S. Takeno and M. Peyrard, Physica D {\bf 92}, 142 (1996); Phys. Rev. E {\bf 55}, 1922 (1997).
\bibitem{Wojciechowski} K. W. Wojciechowski, K. V. Tretiakov, M. Kowalik, Phys. Rev. E {\bf 67}, 036121 (2003).
\bibitem{12} J. Cuevas J. F. R. Archilla, Yu. B. Gaididei and F. R. Romero, Physica D {\bf 163}, 106 (2002).
\bibitem{13} Y. B. Gaididei, S. F. Mingaleev and P. L. Christiansen, Phys Rev. E {\bf 62}, R53 (2000).
\bibitem{14} P. G. Kevrekidis, S. V. Dmitriev, S. Takeno, A. R. Bishop and E. C. Aifantis, Phys. Rev. E (in press).
\bibitem{comp} P. G. Kevrekidis and V. V. Konotop, Phys. Rev. E {\bf 65}, 066614 (2002);
P. G. Kevrekidis, V. V. Konotop and S. Takeno, Phys. Lett. A {\bf 299}, 166 (2002).
\bibitem{15} S. A. Wells, M. T. Dove, M. G. Tucker, K. Trachenko, J. Phys.: Condens. Matter {\bf 14}, 4645 (2002).
\bibitem{16} M. Vallade, B. Berge, G. Dolino, J. Physique I  {\bf 2}, 1481 (1992).
\bibitem{17} S. V. Dmitriev, D. A. Semagin, T. Shigenari, K. Abe, M. Nagamine and T. A. Aslanyan, Phys. Rev. B {\bf 68}, 052101 (2003);
D. A. Semagin, S. V. Dmitriev, K. Abe and T. Shigenari, Russ. J. Phys. Ch. {\bf 77}, S30 (2003).
S. V. Dmitriev, A. A. Vasiliev and N. Yoshikawa, in {\em Recent Research Developments in Physics},
S. G. Pandalai (ed.), (Transworld Research Network, Kerala, India, 2003), Vol. 4, Part I, pp. 267-286.
\bibitem{18} H. Kimizuka, H. Kaburaki, Y. Kogure Phys. Rev. Lett. {\bf 84}, 5548 (2000).
\bibitem{19} M. B. Smirnov, A. P. Mirgorodsky, Phys. Rev. Lett. {\bf 78}, 2413 (1997).
\bibitem{20} N. R. Keskar, J. R. Chelikowsky, Phys. Rev. B {\bf 48}, 16227 (1993).
\bibitem{21} A. Alderson, K. E. Evans, Phys. Rev. Lett. {\bf 89}, 225503 (2002).
\bibitem{22} M. B. Smirnov, Phys. Rev. B {\bf 59}, 4036 (1999).
\bibitem{23} O. M. Braun, Y. S. Kivshar, {\em The Frenkel-Kontorova Model: Concepts, Methods, and Applications},
 (Springer-Verlag, Berlin, 2004).

\end{references}
\end{document}